\documentclass[useAMS,usenatbib]{mn2e}

\usepackage{epsfig}
\usepackage{longtable}
\usepackage{deluxetable}
\usepackage{times}

\title{Testing the $E_{\rm p,i}$-$L_{\rm p,iso}$-$T_{0.45}$ correlation on a {\it BeppoSAX}
and {\it Swift}  sample of gamma--ray bursts}
\author[F. Rossi et al.]{F. Rossi $^{1}$, C. Guidorzi $^{2,3}$, L. Amati $^{4}$,
F. Frontera $^{1,4}$, P. Romano $^{2,3}$, S. Campana $^{3}$, \newauthor
G. Chincarini $^{2,3}$, E. Montanari $^{1,5}$, A. Moretti $^{3}$, G. Tagliaferri $^{3}$\\
$^{1}$Dipartimento di Fisica, Universit\`a di Ferrara, via Saragat 1,
I-44100 Ferrara, Italy\\
$^{2}$Dipartimento di Fisica, Universit\`a di Milano Bicocca,
Piazza delle Scienze 3, I-20126, MIlano, Italy\\
$^{3}$INAF, Osservatorio Astronomico di Brera, via Bianchi, 46,
I-23807 Merate (LC), Italy \\
$^{4}$Istituto Astrofisica Spaziale e Fisica Cosmica, section of Bologna,
CNR/INAF, Via Gobetti 101, I-40129 Bologna, Italy\\
$^{5}$I. I. S. ``I. Calvi'', Finale Emilia (MO), Italy\\
}

\begin{document}

\date{}

\pagerange{\pageref{firstpage}--\pageref{lastpage}} \pubyear{2007}

\maketitle

\label{firstpage}

\begin{abstract}
Using a sample of 14 {\it BeppoSAX} and 74 {\it Swift} GRBs with measured redshift
we tested the correlation between the intrinsic peak energy of the time-integrated
spectrum, $E_{\rm p,i}$, 
the isotropic-equivalent peak luminosity, $L_{\rm p,iso}$, and the duration of the
most intense parts of the GRB computed as $T_{0.45}$ (``Firmani correlation'').
For 41 out of 88 GRBs we could estimate all of the three required properties.
Apart from 980425, which appears to be a definite outlier and notoriously peculiar in
many respects, we used 40 GRBs to
fit the correlation with the maximum likelihood method discussed by D'Agostini,
suitable to account for the extrinsic scatter in addition to the intrinsic
uncertainties affecting every single GRB.
We confirm the correlation. However, unlike the results by Firmani et al., we found 
that the correlation does have a logarithmic scatter comparable with that of the
$E_{\rm p,i}$-$E_{\rm iso}$ (``Amati'') correlation.
We also find that the slope of the product $L_{\rm p,iso}\,T_{0.45}$ is equal
to $\sim0.5$, which is consistent with the hypothesis that the
$E_{\rm p,i}$-$L_{\rm p,iso}$-$T_{0.45}$ correlation is equivalent to the
$E_{\rm p,i}$-$E_{\rm iso}$ correlation (slope $\sim0.5$).
We conclude that, based on presently available data, there is no clear evidence that the
$E_{\rm p,i}$-$L_{\rm p,iso}$-$T_{0.45}$ correlation is different (both in terms of
slope and dispersion) from the $E_{\rm p,i}$-$E_{\rm iso}$ correlation.
\end{abstract}


\section{Introduction}
Ten years after the first measurements of the cosmological distances to Gamma-Ray Bursts
(GRBs) made possible by the {\it BeppoSAX} satellite \citep{Boella97}, the task
of measuring the redshift either of their afterglow itself or of the host galaxy
associated with a GRB remains challenging.
In the era of the {\it Swift} spacecraft \citep{Gehrels04} the rate of
GRBs with a measured distance has increased remarkably thanks to its rapid follow-up
capabilities and its arcsec X-ray localisations promptly distributed for most
GRBs. Yet, only for one third out of over $\sim$300 GRBs detected to date (April 2008)
the redshift measurement is available. In the remaining cases, due to the combination of
unfavourable observing conditions, such as high Galactic extinction or intrinsic
faintness of the afterglow or unavailability of equipped telescopes especially
with the NIR filters for high-$z$ GRBs, the attempt is doomed to failure
(e.g., see Fynbo et al., 2007).\nocite{Fynbo07}

With respect to the long duration GRBs with known redshift,
several correlations between intrinsic properties have already been discovered
(e.g. see Schaefer \& Collazzi 2007\nocite{Schaefer07}). The interest in these correlations is
twofold: they are a direct way to test the predictions of the emission mechanisms
models and, in perspective, they could potentially be used as luminosity estimators.
Among the most popular and debated examples, we mention the relation discovered by
\citet{Amati02} between the rest-frame peak energy of the 
high-energy $\nu F_{\nu}$ spectrum of the prompt emission, $E_{\rm p,i}$, and
the isotropic-equivalent radiated energy in the rest-frame 1--10000~keV energy band,
$E_{\rm iso}$, that shows a dispersion of the data points with 
$\sigma_{\log{E_{\rm p,i}}}=0.15\pm0.04$~\citep{Amati06}.
A tighter correlation has been found afterwards between $E_{\rm p,i}$ and
the collimation-corrected radiated energy, $E_{\gamma}$, where the jet angle
is derived from the time of the break in the afterglow light curve \citep{Ghirlanda04}.
Both relations are still a matter of debate. Criticisms to the
$E_{\rm p,i}$-$E_{\rm iso}$ relation have been raised by \citet{NP05} and \citet{Band05}, 
who claim that the majority of GRBs with unknown redshift detected with {\it CGRO}/BATSE
\citep{Paciesas99} are inconsistent with this relation.
However, different results have been reported by other
authors \citep{Ghirlanda05b,Pizzichini06}. See \citet{Amati06} for an updated 
review on this subject.
In the case of the $E_{\rm p,i}$-$E_{\gamma}$, what appears to be a crucial
and often controversial task is the identification of the break (if any) in the
afterglow light curve due to the jet \citep{Panaitescu06,Ghirlanda07,Campana07}.
A similar and less model-dependent relation has been found between $E_{\rm p,i}$,
$E_{\rm iso}$ and the rest-frame break time of the optical afterglow light curve
$t_{\rm b}$ \citep{LZ05}.

Other correlations have been reported in the literature, such as that between the temporal
variability of the time profile and the peak luminosity
\citep{Fenimore00,R01,Schaefer01}. However, recent work based on larger samples
proved that the dispersion of this relation is so large as to make it a useless
luminosity estimator \citep{CG05,CG05batse,Rizzuto07}.

The correlation found by \citet{Norris00} between peak luminosity and spectral lag
(estimated by cross-correlating time profiles of the same GRB at different energy bands)
appears to be a promising tool for identifying the short duration GRBs characterised
by an initial spike, followed by a soft and long tail, which otherwise may look like
long GRBs \citep{Norris06}.

In this paper we focus on the correlation discovered by \citet{Firmani06}
involving three properties of the GRB prompt emission: $E_{\rm p,i}$, the
isotropic-equivalent peak luminosity in the rest-frame 1--10000~keV band,
$L_{\rm p,iso}$, and the smoothing timescale $T_{0.45}$ as defined by \citet{R01}.
This correlation ($L_{\rm p,iso} \propto E_{\rm p,i}^{1.62}\,T_{0.45}^{-0.49}$)
was derived from a sample of 22 GRBs and was found to be very tight.
This feature would make it an ideal luminosity estimator.
In fact, if one assumes $E_{\rm iso}\propto L_{\rm p,iso}\,T_{0.45}$ and the validity
of the $E_{\rm p,i}$-$E_{\rm iso}$ relation, the above correlation follows
straightforwardly. 

We test the $E_{\rm p,i}$-$L_{{\rm p,iso}}$-$T_{0.45}$
correlation using a larger sample (88) of GRBs with known redshift 
from {\it BeppoSAX} and {\it Swift}.
In particular, we study $E_{\rm p,i}$ as a function of $L_{\rm p,iso}$ and $T_{0.45}$
to compare its dispersion with that of the $E_{\rm p,i}$-$E_{\rm iso}$ relation.
In Section~\ref{s:sample} we present our sample of GRBs; in Section~\ref{s:method}
we illustrate the data analysis. In Section~\ref{s:results} we present and discuss
our results.

\section[]{The GRB sample}
\label{s:sample}
The sample of 88 long GRBs with known redshift includes 14 GRBs detected by the
Gamma-Ray Burst Monitor (GRBM; Feroci et al., 1997; Frontera et al., 1997;
Costa et al., 1998)\nocite{Feroci97,Frontera97,Costa98} aboard 
{\it BeppoSAX} and 74 by the Burst Alert Telescope (BAT; Barthelmy,
2005)\nocite{Barthelmy05} aboard {\it Swift}. For
the latter we consider the GRBs from January 2005 to April 2008.
Table~1 reports the full list of GRBs of our sample.

The shortest time binning available for the {\it BeppoSAX}/GRBM data was 7.8125~ms 
in the 40--700~keV energy band. For the {\it Swift}/BAT data the time binning was 
set to 64 ms in order to ensure a good signal-to-noise (SNR) ratio.

For the GRBM data we considered all the GRBs with known redshift that have a firm 
estimate of $E_{\rm p,i}$ as reported in \cite{Amati06} and for which high 
resolution data were acquired (because of this, we excluded 980613, 011211). For 
990510 we used the BATSE data with 64 ms time binning. For 000210 we considered the 
light curve as in \cite{CG05}.

As far as BAT GRBs are regarded, we selected only the events whose $\gamma$-ray 
profile is entirely covered by BAT during the burst mode. Due to these selection 
criteria we rejected 050318, 050820A, 050904, 060218 and 060906. GRB~060124 was not 
included in the sample because only the precursor was recorded in burst mode 
\citep{Romano06}, while the main event was covered by the survey mode, whose coarse 
time resolution makes it unsuitable to our aim.

\section[]{Data analysis}
\label{s:method}

\subsection{T$_{0.45}$}

The smoothing timescale $T_{f}$, defined by \cite{R01}
for the calculation of the variability, is the shortest cumulative time interval
covering the 100$f$\% of the total counts above the background.
The fraction $f$ was set to $0.45$ because it was originally found to maximise
the correlation between variability and peak luminosity \citep{R01}.
A correct evaluation of $T_{0.45}$ must fulfil the requirements found by \cite{CG05}.
For the BAT and GRBM GRBs already published, the values of $T_{0.45}$ are consistent
with those reported by \cite{Rizzuto07} and \cite{CG05}, respectively.
The $T_{0.45}$ of GRB~980703 reported in this paper differs from that reported
by \cite{CG05} and is consistent with that obtained by \cite{R01} on BATSE data.
However, we verified that this had a negligible impact on the variability estimate
obtained by \cite{CG05} for this specific GRB.

\subsubsection{$T_{0.45}$ as function of energy}

In our sample we have two data sets, one for GRBM events in the 40--700 keV band, 
the other for BAT events in the 15--150 keV band. Given that the value
of $T_{0.45}$ is dependent on the energy band used, we modelled
this dependence with a power law, similarly to what originally done by \citet{Fenimore95},
for the energy dependence of the autocorrelation function width.

\begin{figure}
\begin{center}
\centerline{\includegraphics[scale=0.7]{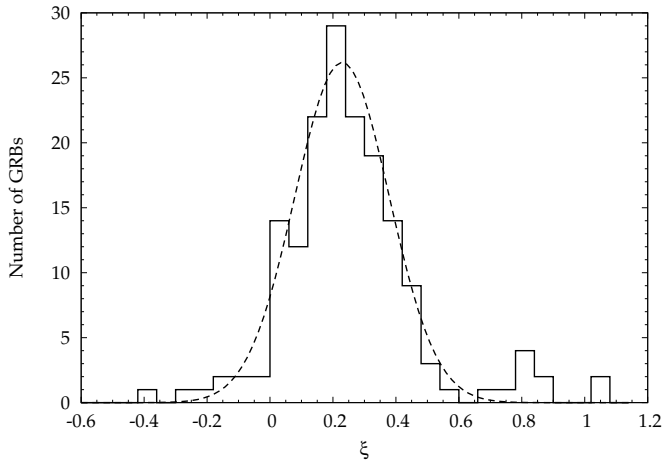}}
\caption{Distribution of the power--law index $\xi$ modelling the dependence of
$T_{0.45}$ on energy.
The dashed line shows the best-fitting normal distribution centred on $\overline{\xi} =0.23$
and $\sigma = 0.15$.}
\label{f:distribution_csi}
\end{center}
\end{figure}

We considered 284 \textit{{\it Swift}}/BAT GRBs (all the \textit{{\it Swift}}/BAT
GRBs from January 2005 to April 2008 regardless of the redshift availability)
and for each of them we calculated $T_{0.45}$ in the four nominal BAT energy channels:
15--25, 25--50, 50--100 and 100--150~keV.
We focused on the GRBs with an accurate value of $T_{0.45}$ in, at least, three
BAT energy channels. As a consequence, only 164 GRBs were selected.
For each of them we modelled $T_{0.45}$ with a power law:
$T_{0.45}(E) \propto E^{-\xi}$.
Figure~\ref{f:distribution_csi} shows the distribution of the power-law index $\xi$:
this is consistent with being normally distributed around the mean value of
$0.23$ and $\sigma =0.15$.
We note that this dependence on energy is marginally less strong than that 
obtained for the autocorrelation function width by \nocite{Fenimore95}Fenimore et al. (2005; 
power-law index of 0.4) with BATSE and fully consistent with the energy 
dependence of the autocorrelation function width found by \citet{Borgonovo07} 
for a sample of 19 {\it BeppoSAX} GRBs. This is not surprising, given that the BAT
energy band (15--150 keV) is somewhat between that of the {\it BeppoSAX}/WFC+GRBM (2--700~keV) 
and that of BATSE ($>25$~keV).

%
In order to establish the best reference energy $E_{\rm r}$ at which we have to estimate 
$T_{0.45}$ in the rest frame (obtained dividing the observed $T_{0.45,{\rm obs}}$ by $(1+z)$), 
for all GRBs in our sample we adopted  the following approach.
Given that  $T_{{0.45},{\rm obs}}$ of each GRB is mostly dominated by photons 
with energies close to their mean energy, for each GRB $i$ in our sample we first
determined its rest-frame mean energy $E_{m,i}$. 
The best rest-frame energy $E_{\rm r}$ was obtained by performing
a logarithmic mean of the derived $E_{m,i}$, finding for the entire sample 
a value $E_{\rm r}=145$~keV.
The values of $T_{0.45}$ at this energy are reported in Table~1.

\subsection{$E_{\rm p,i}$}
For each GRB in our sample we evaluated the rest-frame peak energy 
$E_{\rm p,i}$ of the $E\,F(E)$ time averaged spectrum.
For the {\it BeppoSAX} GRBs we considered the approach followed by
\cite{Amati02}, i.e., we fitted the spectra with a smoothly joined
power-law proposed by \cite{Band93}, whose parameters, in addition
to the normalisation,  are the break energy $E_0$, and
the low and high energy indices $\alpha$ and $\beta$, respectively.

For the \textit{Swift}/BAT GRBs, to obtain a firm estimate
of $E_{\rm p,i}$ the above approach was not always possible because of the 
relatively narrow BAT energy band. Therefore, when available,
we adopted the $E_{\rm p,i}$ values obtained for the same GRBs with
the {\it Konus}/WIND experiment. In the other cases, we used the
values derived from the BAT spectra averaged over the
$T_{90}$ interval. The estimated values or their upper/lower limits are generally
quite consistent with those reported by \citet{Butler07} and \citet{Sakamoto07}.

We also checked other integration times of the BAT spectra, e.g., time intervals 
based on a significance threshold with respect to the background for each GRB,
finding photon indices slightly softer than those reported in 
Table~2, but the statistical quality of the spectra was worse.

\subsection{L$_{p,{\rm iso}}$}
The GRB peak luminosity, $L_{p, {\rm iso}}$, 
in the source cosmological rest--frame 1--10000 keV energy band is given by:

\begin{equation} 
L_{p,{\rm iso}}=4\pi D_{L}^{2}(z)\int_{1/(1+z)}^{10000/(1+z)}E\,\Phi(E)\,dE 
\end{equation} 
where $\Phi(E)$ is the measured spectrum at the peak 
(ph~cm$^{-2}$~s$^{-1}$~keV$^{-1}$), $D_{L}$ is the luminosity distance at 
redshift $z$ (using $H_0 = 71$~km~s$^{-1}$~Mpc$^{-1}$, $\Omega_M = 0.27$ and 
$\Omega_{\Lambda}=0.73$) and $E$ is the energy expressed in keV. 

Depending on the morphology of the 
main pulse (e.g. smooth or spiky), the peak flux, and thus $L_{p, {\rm iso}}$,
may vary up to a factor 
$1.5$--$2$ when different time scales are considered for its computation (e.g., from 
the commonly used 1~s to 64~ms time scale).
It must also be noted that a 
fixed time scale in the observed light curve corresponds to different 
rest--frame time scales 
for GRBs at different redshifts, thus providing peak luminosities computed in an
inhomogeneous way. In addition, the spectral shape at the peak is often uncertain, 
because of the low statistical quality of the data and/or the limited number 
of channels and integration times for time resolved spectra provided by 
instruments. This is particularly true for {\em Swift}/BAT, because of its narrow 
energy band, and for {\em BeppoSAX}/GRBM, which provided time resolved spectra only in 2 
channels and with a time resolution of 1~s.

We computed the peak luminosities of the GRBs included 
in our sample by following three different methods. First of all, in 
order to perform a comparison between our and their results, we followed the same 
method used by \citet{Firmani06}. We extracted the peak spectrum integrated 
over 1~s and fit it with a Band model in which $\alpha$, $\beta$ and 
$E_0$ were frozen to the best-fitting values obtained for
the spectrum averaged on the entire GRB time profile, while the 
normalisation was left free to vary. This method is 
the most commonly used in the literature, e.g., in works studying the $L_{\rm 
p,iso}$--$E_{\rm p,i}$ correlation \citep{Ghirlanda05a,Yonetoku04}.
We call this ``1~s'' time scale method.

In the second case, we computed $L_{{\rm p,iso}}$ using for each GRB the spectrum
integrated over the shortest time interval around its peak so as to have a significant number
of counts for each energy channel. The spectrum was then fit by still freezing $\alpha$,
$\beta$ and $E_0$ to the corresponding values of the time-averaged spectrum, as done
for the previous method. We call this "variable" time scale method.

We also attempted to evaluate $L_{\rm p,iso}$ by fitting the spectrum integrated
over the variable time scale, as above, with all the spectral parameters left free to vary. 
In principle this should be the best method for the peak luminosity estimate.
However, the statistical quality of these spectra, their narrow energy passband
(in the case of BAT), the low number of channel spectra above 30 keV (in the case of GRBM) 
did not allow to get well constrained estimates of the GRB peak luminosity.

In the following text, in the Tables and in the Figures we call the peak 
luminosities computed with the two different methods described above as
$L_{{\rm p,1s}}$ and $L_{\rm p,var}$, respectively.
The results of both methods are reported in Table~1.

\begin{figure}
\begin{center}
\centerline{\includegraphics[scale=0.7]{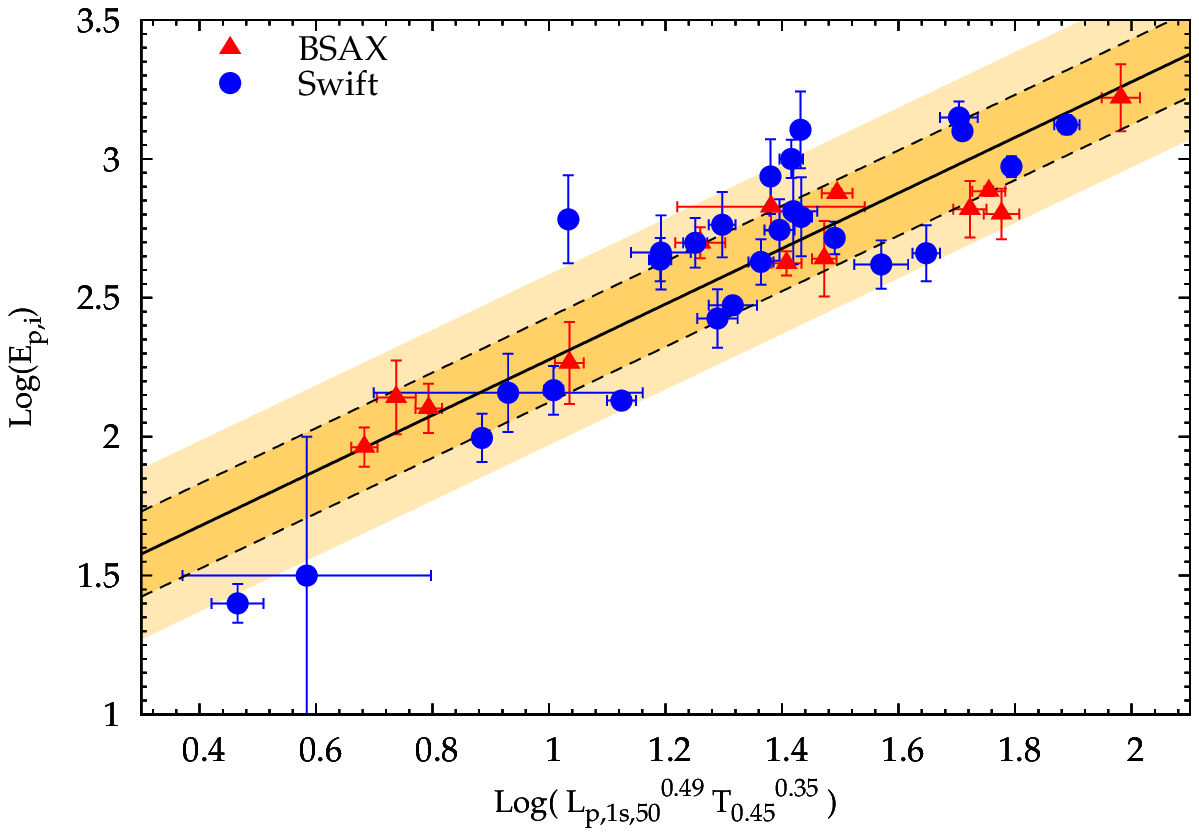}} 
\centerline{\includegraphics[scale=0.7]{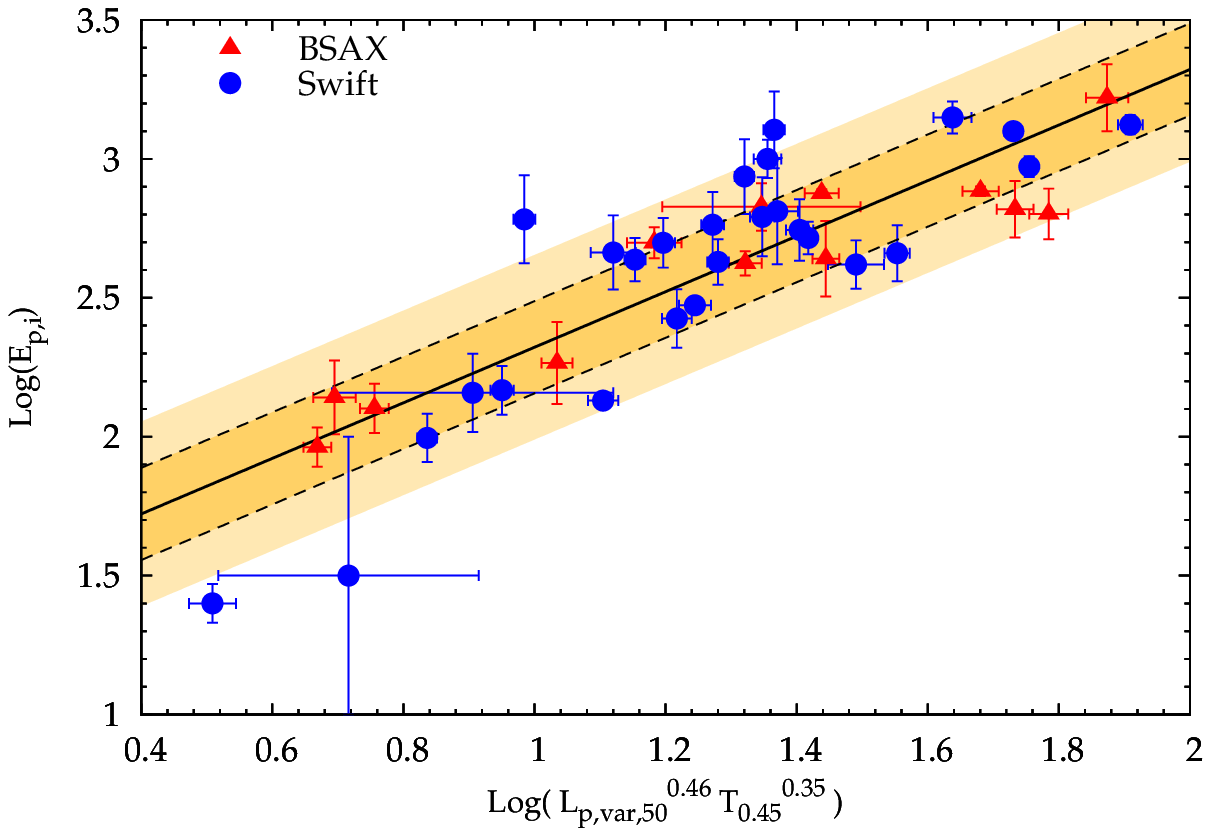}} 
\caption{Peak energy $E_{\rm p,i}$ as a function of $L_{\rm p,iso}$ and $T_{0.45}$.
Only GRBs with firm estimates of all of the properties are shown. Triangles (circles)
correspond to {\it BeppoSAX} ({\it Swift}) GRBs. Peak luminosities are expressed
in units of $10^{50}$~erg~s$^{-1}$.
In the top plot the peak luminosity used is $L_{\rm p,1s}$, in the bottom it is 
$L_{\rm p,var}$ (see text). The solid line shows the best fit curve obtained
with the maximum likelihood method illustrated by \citet{Dagostini05}.
Dashed lines represent the 1-$\sigma$ region. Shaded areas show the 1- and 2-$\sigma$ regions.
$\sigma$ is the best-fit value found for $\sigma_{\log{E_{\rm p,i}}}$ in each case.}
\label{f:Ep-Lp-T045-no-lim}
\end{center}
\end{figure}
%
\section{Results}
\label{s:results}
In order to study the dependence of $E_{\rm p,i}$ on 
both $L_{\rm p,iso}$ and $T_{0.45}$, we first used the 40 GRBs in our sample (see
Table~1) for which we have a firm determination of
of $E_{\rm p,i}$, $L_{\rm p,iso}$ and $T_{0.45}$.
We applied the maximum likelihood method (hereafter MLM) discussed by 
\citet{Dagostini05} extended to three variables. This method, already
adopted by us for other correlation studies (see Guidorzi et~al. 2006; Amati 2006),
\nocite{Guidorzi06,Amati06} is the best tool to take into account,
in addition to the statistical uncertainty in the parameters,
the so called extrinsic (or external) scatter,
that is the scatter due to the presence of unknown variables that influence the
correlation to be tested.

We modelled the correlation among $E_{\rm p,i}$, $L_{\rm p,iso}$ and $T_{0.45}$, 
according to the equation 
\begin{equation}
\log{(E_{\rm p,i})} = a\,\log{(L_{\rm p,iso})} + b\,\log{(T_{0.45})} + q
\label{e:corr}
\end{equation}
where the four parameters $a$, $b$, $q$ and the extrinsic scatter $\sigma_{\log{E_{\rm p,i}}}$
are free to vary in the fit.
The best-fitting parameters so obtained, for each of the two peak luminosity
estimates, are reported in Table~\ref{t:parametri_ep_LT}, 
together with their uncertainties (at 90\% confidence level), and the best fit $\chi^2$ and
chance probability.
 
As can be seen from this Table, for both $L_{\rm p,1s}$ and $L_{\rm p,var}$ 
estimates of $L_{\rm p,iso}$, we find a significant value of extrinsic scatter,
as displayed in Fig.~\ref{f:Ep-Lp-T045-no-lim}.
This result is confirmed by the fit of the data with Eq.~\ref{e:corr}, freezing 
$\sigma_{\log{E_{\rm p,i}}}$ to 0. The resulting fit, also reported in 
Table~\ref{t:parametri_ep_LT}, is highly unacceptable.

GRB~980425 was found not to follow at all the correlation, as in the case of the other 
relations, such as the $E_{\rm p,i}$--$E_{\rm iso}$ \citep{Amati02}, the lag--luminosity
\citep{Norris00} and the variability--luminosity \citep{R01} ones.
This GRB is also peculiar for several aspects, such as its being subluminous and
its association with SN1998bw, thanks to which it was possible to measure its relatively
close ($\sim40$~Mpc) distance. 
Therefore, like \citet{Firmani06}, we did not include it in the sample
used to derive the fit results in Table~\ref{t:parametri_ep_LT}, and we focused on the
canonical long-duration GRBs and XRFs which are known to follow all of the main correlations.
If we include GRB~980425 in the sample, its contribution to the dispersion of the correlation
is remarkable: in the $L_{\rm p,1s}$ case, the extrinsic scatter passes
from $0.15_{-0.03}^{+0.05}$ to $0.22\pm0.05$ (the other parameters becoming $a=0.32\pm0.06$,
$b=0.29\pm0.16$ and $q=1.73\pm0.19$, respectively).
Similar results are obtained using $L_{\rm p,var}$.

Moreover, we applied the MLM also to the subsample of 27 {\em Swift} GRBs
with determined $E_{\rm p,i}$, $T_{0.45}$ and $L_{\rm p,iso}$.
The results, reported in Table~2, clearly show that also in this case
the extrinsic scatter, $\sigma_{\log{E_{\rm p,i}}}=0.17\pm0.04$ ($L_{\rm p,1s}$), is
fully consistent with that derived from the entire sample and far from being negligible.
This proves that the extrinsic scatter is a property of the correlation itself and not
an artifact of merging data sets from different instruments.

\begin{figure*} 
\begin{center} 
\centerline{\includegraphics[width=15cm]{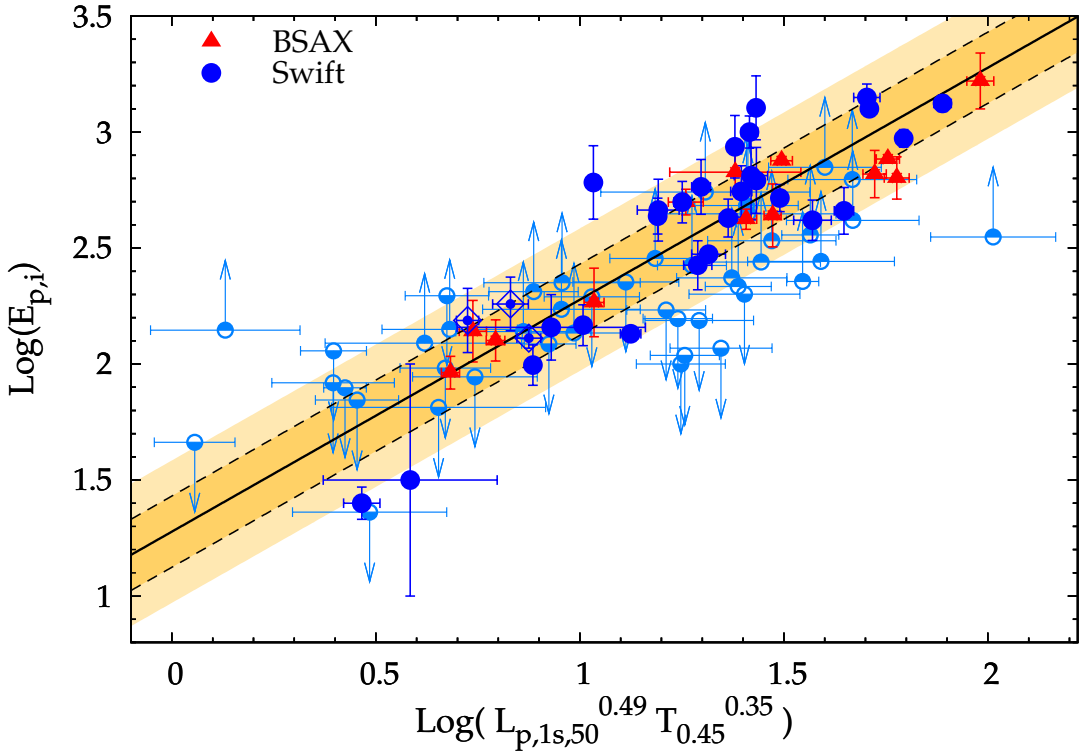}} 
\caption{Peak energy $E_{\rm p,i}$ as a function of $L_{\rm p,1s}$ and $T_{0.45}$ 
using all GRBs in our sample. Triangles (circles) correspond to {\it BeppoSAX} 
({\it Swift}) GRBs. Dashed lines represent the 1-$\sigma$ region. Shaded areas 
show the 1- and 2-$\sigma$ regions.
$\sigma$ is the best-fit value found for $\sigma_{\log{E_{\rm p,i}}}$.
The empty diamonds show the three {\em Swift}
GRBs, 070506, 070611 and 070810A, for which we constrained $E_{\rm p,i}$ but
which were not used to fit correlation (see text).}
\label{f:Ep-Lp2-T045-lim} 
\end{center} 
\end{figure*}

Including all GRBs in our sample, the data points are shown in  Fig.~\ref{f:Ep-Lp2-T045-lim}, 
where we report also the best fit relation between
$E_{\rm p,i}$, $L_{\rm p}$  and $T_{0.45}$ in the case
of the $L_{\rm p,1s}$ estimate and with the extrinsic scatter taken into 
account.
As can be seen from this figure, in addition to many lower or upper limits to
$E_{\rm p,i}$ potentially consistent with the correlation, a few of them clearly
deviate by more than 2$\sigma$ from the best fit model.

\begin{figure} 
\begin{center} 
\centerline{\includegraphics[scale=0.7]{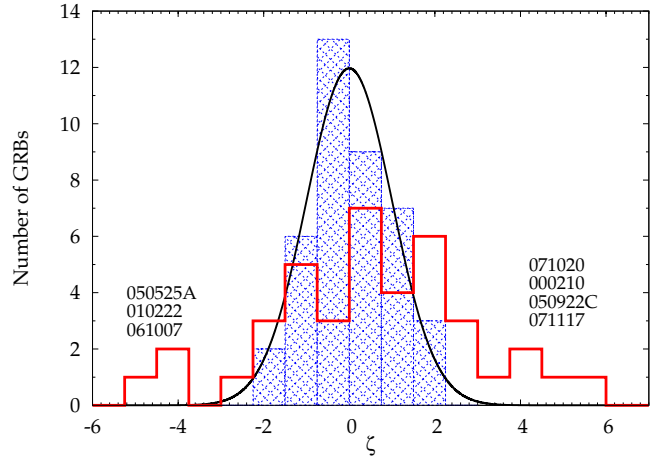}} 
\caption{The shaded (thick line) histogram shows the distribution of the 
normalised scatter $\zeta$ (not) inclusive of the extrinsic scatter. The expected 
normal $N(0,1)$ is also displayed. The peak luminosity used
is $L_{\rm p,1s}$.} 
\label{f:scatter_distrib} 
\end{center} 
\end{figure}

In order to understand the origin the extrinsic scatter, we studied the 
distribution $N(\zeta)$ of the normalised deviation of the 
measured $\log{E_{\rm p,i}}$ from the values expected on the basis of the best fit curve
(Eq.~\ref{e:corr}) in two cases, i.e., by including or excluding the found extrinsic scatter 
$\sigma_{\log{E_{\rm p,i}}}$:
\begin{equation}
\zeta_{i} = \frac{(a\,\log{L_{\rm p,iso}^{(i)}}+b\,\log{T_{0.45}^{(i)}}+q) - \log{E_{\rm p,i}^{(i)}}}{\sqrt{\sigma_{\log{E_{\rm p,i}^{(i)}}}^2+a^2\,\sigma_{\log{L_{\rm p,iso}^{(i)}}}^2+
b^2\,\sigma_{\log{T_{0.45}^{(i)}}}^2 + \sigma_{\log{E_{\rm p,i}}}^2}}
\label{e:norm_scatter}
\end{equation}
Figure~\ref{f:scatter_distrib} shows the result in the case of the peak luminosity estimate 
$L_{\rm p,1s}$. When the extrinsic scatter is taken into account, the resulting 
distribution (shaded histogram) is consistent with a normalised Gaussian, consistently with 
the picture of an extrinsic scatter characterising the correlation itself. 
Instead, assuming no extrinsic scatter ($\sigma_{\log{E_{\rm p,i}}}=0$),
we find an histogram (see the thick line in Fig.~\ref{f:scatter_distrib}) clearly 
inconsistent with the normalised Gaussian. In addition to seven apparent outliers 
lying $>$\,$3$~$\sigma$ off (071020, $+5.5$~$\sigma$; 000210, $+4.4$~$\sigma$;
071117, $+3.9$~$\sigma$; 050922C, $+3.8$~$\sigma$; 050525A, $-5.2$~$\sigma$;
061007, $-4.0$~$\sigma$; 010222, $-4.0$~$\sigma$), others
GRBs contribute to broaden the histogram, making it inconsistent with a normalised Gaussian.

We have carefully checked the adopted estimates of $E_{\rm p,i}$, $L_{\rm p,iso}$ 
and $T_{0.45}$ attributed to the outliers, finding that have they are robust. 
For example, in the case of GRB~000210 (a {\it BeppoSAX} GRB) we confirm the correctness 
of the attributed values. In the case of GRB~050525A, the accurate estimate of
 $E_{\rm p,i}$ was provided by the {\it Konus}/WIND experiment and we see no reason 
to reject it.


Similar results are obtained when the same analysis is performed using $L_{\rm p,var}$
as peak luminosity estimate.

\section{Discussion}
\label{s:disc}

After the discovery by \citet{Firmani06} of a correlation among
the rest frame quantities $E_{\rm p,i}$, $L_{\rm p,iso}$, and $T_{0.45}$, 
obtained with a sample of 22 GRBs ('Firmani' relation), using a larger GRB sample 
(88 GRBs with known redshift detected by {\it BeppoSAX} and {\it Swift})
the correlation has been re-tested. 
By ignoring the outlier GRB~980425 and 47 GRBs for which only upper/lower limits 
to $E_{\rm p,i}$ were possible to be established, 
we confirm the correlation with only slightly different best-fitting parameters
(see Table~\ref{t:parametri_ep_LT}). However, unlike \citet{Firmani06} we find a 
%
%
a significant extrinsic dispersion of the data points around the best fit curve,
parametrised by the $\sigma_{\log{E_{\rm p,i}}}$ value reported in 
Table~\ref{t:parametri_ep_LT}, that denotes 
the presence of an unknown variable (see D'Agostini 2005). This scatter
is found to be independent of the time integration of the measured spectra 
(either 1~s for all GRBs or variable from a GRB to another depending on the 
light curve shape and statistical quality) used to estimate of $L_{\rm p,iso}$.

It is also apparent from the $\chi^2/{\rm dof}$ reported Table~\ref{t:parametri_ep_LT}, 
that assuming a null extrinsic scatter gives unacceptable results.
We have analysed the origin of the extrinsic scatter and found that, in addition to
seven clear outliers, other GRBs contribute to the found dispersion.
From a detailed analysis of the data available for each of the 40
GRB included in our reduced sample, we cannot find any reason to infer that 
some of the estimates reported in Table~2 is unreliable.

Furthermore, as can be seen in Fig.~\ref{f:Ep-Lp2-T045-lim},
also some GRBs with only upper/lower limits deviate form the best fit curve
by more than 2~$\sigma$. 
Thus, the distribution of GRBs in the correlation plane found by us is not as
tight as that found by \citet{Firmani06} using a smaller sample of 22 events.

On the basis of the reported results, we derive an interesting consequence. 
Taking into account that, in the $E_{\rm p,i}$, $L_{\rm p,iso}$, $T_{0.45}$ 
multivariate correlation, the best-fitting power-law  indices
for $L_{\rm p,iso}$ and  $T_{0.45}$ (see Table~~\ref{t:parametri_ep_LT}) 
are both consistent with $0.5$, we infer that the Firmani relation can be
approximately written as 
\begin{equation} 
E_{\rm p,i} \propto (L_{p,1s} T_{0.45})^{0.5} 
\end{equation}

That renders the Firmani relation equivalent to the $E_{\rm p,i}$ vs. $E_{\rm iso}$ 
relation discovered by \citet{Amati02}. Also the obtained extrinsic scatter is consistent
with that of the Amati relation \citep{Amati06}.
In conclusion, it seems that the Firmani relation does not provide 
more information than that contained in the Amati relation.
It is expected that the future joint observations by {\em Swift}
and {\em GLAST} will provide a sizable set of GRBs with firm measures of all the required
observables, thus allowing to refine the estimate of the dispersion
and to better characterise its link with the $E_{\rm p,i}$--$E_{\rm iso}$ relation.

%
%
\onecolumn
\begin{deluxetable}{lccccrcrrccc}
\label{tab:total_table}
\tabletypesize{\small}
\rotate
\tablecaption{The GRB sample: $T_{f=0.45}$ (at the rest-frame energy of $E_{\rm 
r}=145$~keV), the intrinsic peak energy, $E_{\rm p,i}$, and the peak luminosities 
$L_{\rm p,1s}$ and $L_{\rm p,var}$.}
\tablewidth{0pt}
\tablehead{
\colhead{GRB} & \colhead{z} & \colhead{Inst.$^{\rm (a)}$} & 
\colhead{$T_{f=0.45}$} & \colhead{$E_{\rm p,i}$} &
\colhead{ $t_{\rm start,1s}^{\rm (b)}$} & \colhead{$L_{\rm p,1s}^{\rm (c)}$} &
\colhead{ $t_{\rm start,var}^{\rm (b)}$} & \colhead{$t_{\rm stop,var}^{\rm (b)}$} &
\colhead{$L_{\rm p,var}^{\rm (c)}$} &
\colhead{$z$} & \colhead{$E_{\rm p,i}$} \\ 
\colhead{Name} & \colhead{Redshift} &  &
\colhead{(s)} & \colhead{(keV)} & 
 \colhead{(s)} & \colhead{ ($10^{50}$~erg s$^{-1}$)} &
\colhead{(s)} & \colhead{(s)} &
\colhead{ ($10^{50}$~erg s$^{-1}$)} &
\colhead{Reference $^{\rm d}$} & \colhead{Reference $^{\rm e}$}\\
\colhead{1} & \colhead{2} & \colhead{3} & \colhead{4} & \colhead{5} & \colhead{6} & \colhead{7} & \colhead{8} & \colhead{9} & \colhead{10} & \colhead{11}& \colhead{12}
}
\startdata
970228$^{\rm f}$  & $0.695$ & G & $1.54_{-0.19}^{+0.18}$ & $195\pm64$  &  $0.437$ & $95\pm7$ &  $0.680$ &  $0.774$ &  $130\pm10$  &  1 & 1 \\
970508$^{\rm f}$  & $0.835$ & G & $1.70_{-0.29}^{+0.31}$ & $145\pm43$  &  $0.133$ & $22\pm2$ &  $0.196$ &  $1.132$ &  $22\pm2$    &  2 & 1 \\
971214$^{\rm f}$  & $3.42$  & G & $1.45_{-0.31}^{+0.39}$ & $680\pm130$ & $10.857$ &$650\pm400$& $6.068$ &  $6.248$ &  $830\pm520$ &  3 & 1 \\
980425            & $0.0085$& G & $4.46_{-0.25}^{+0.29}$ & $55 \pm21$ & $3.977$ & $(6.6\pm0.6)\times10^{-4}$&$2.797$ &$4.945$ &$(6.0\pm0.6)\times10^{-4}$ &4 & 1 \\
980703$^{\rm f}$  & $0.966$ & G & $11.52_{-2.50}^{+2.68}$& $503\pm64$  &  $2.656$ & $65\pm8$ &  $2.456$ &  $4.456$ &  $60\pm7$   &  5 &  1 \\
990123$^{\rm f}$  & $1.60$  & G & $6.54_{-1.14}^{+1.39}$ & $1720\pm470$&  $6.758$ &$2850\pm200$&  $6.750$ &  $7.172$ &  $2900\pm200$& 6 & 1 \\
990506$^{\rm f}$  & $1.30$  & G & $4.84_{-0.76}^{+0.90}$ & $680\pm160$ & $87.055$ &$1050\pm60$ & $90.180$ & $90.202$ & $1800\pm100$ & 7 & 1 \\
990510$^{\rm f}$  & $1.619$ & G & $1.53_{-0.22}^{+0.24}$ & $423\pm42$  & $40.332$ & $550\pm30$ & $40.332$ & $41.332$ & $550\pm30$  & 8 & 1 \\ 
990705$^{\rm f}$  & $0.86$  & G & $7.94_{-0.95}^{+1.08}$ & $460\pm140$ & $10.514$ & $230\pm10$ & $20.842$ & $20.936$ & $290\pm10$  & 9, 10 & 1\\
990712$^{\rm f}$  & $0.434$ & G & $3.54_{-0.25}^{+0.26}$ & $93\pm15$   &  $0.625$ & $10\pm1$  &  $0.679$ &  $1.046$ & $11\pm1$     & 11 &   1  \\
991216$^{\rm f}$  & $1.02$  & G & $1.67_{-0.24}^{+0.28}$ & $650\pm130$ &  $3.289$ & $2900\pm220$& $4.156$ & $4.172$ & $5200\pm400$ & 12 & 1  \\
000210$^{\rm f}$  & $0.846$ & G & $1.06_{-0.14}^{+0.16}$ & $750\pm30$  &  $3.022$ & $1100\pm80$ & $3.248$ & $3.310$ & $1300\pm100$ & 13 & 1  \\
010222$^{\rm f}$  & $1.477$ & G & $3.42_{-0.53}^{+0.62}$ & $766\pm30$  & $58.680$ & $1600\pm90$ &$58.875$ & $58.953$ & $1800\pm100$ & 14 &  2 \\
010921$^{\rm f}$  & $0.45$  & G & $5.04_{-0.51}^{+0.53}$ & $129\pm26$  & $10.383$ & $13\pm1$  &  $9.586$ & $10.930$ & $13\pm1$  & 15 & 1   \\
050126            & $1.29$  & B & $5.00_{-0.52}^{+0.67}$ & $>172$      & $4.112$  &       -       &  $2.448$ &  $5.008$ & -    & 16 & 3 \\
050223            & $0.5915$& B & $4.98_{-1.00}^{+0.88}$ & $<114$      & $1.584$  &       -       &  $6.256$ & $11.120$ & -    & 17 & 3 \\
050315            & $1.949$ & B & $6.19_{-0.41}^{+0.42}$ & $<109$      & $24.592$ &       -       & $24.720$ & $25.296$ & -    & 18 & 3 \\
050319            & $3.240$ & B & $3.38_{-0.46}^{+0.31}$ & $<157$      & $0.656$  &       -       &  $0.336$ &  $0.976$ & -    & 19 & 3 \\
050401$^{\rm f}$  & $2.90$  & B & $1.43_{-0.13}^{+0.13}$ & $470\pm110$ & $24.248$ & $1780\pm160$ & $24.760$ & $25.272$ & $1850\pm130$ & 20 &1\\
050416A$^{\rm f}$ & $0.6535$& B & $0.70_{-0.14}^{+0.15}$ & $25.1_{-3.7}^{+4.4}$ & $-0.064$ & $11.7\pm1.7$ & $0.704$&$0.896$ & $17.0\pm1.4$& 21&1\\
050505            & $4.27$  & B & $2.72_{-0.43}^{+0.31}$ & $>416$      & $1.000$  &       -       & $1.000$  & $2.024$  & -    & 22 & 3 \\
050525A$^{\rm f}$ & $0.606$ & B & $1.37_{-0.15}^{+0.17}$ & $135\pm3$   & $0.848$  & $157\pm13$    & $1.232$  & $1.360$  & $200\pm15$  & 23 & 1\\
050603$^{\rm f}$  & $2.821$ & B & $0.68\pm0.06$          & $1330\pm110$& $-0.184$ & $9400\pm700$  & $0.136$  & $0.264$  & $19300\pm1200$&24 &1\\
050730            & $3.967$ & B & $12.06_{-1.10}^{+1.12}$& $>705$      & $4.408$  &       -       & $2.488$  & $7.288$  & -    & 25 & 3 \\
050803            & $0.422$ & B & $11.94_{-1.77}^{+1.72}$& $>123$      & $147.208$&       -       & $147.528$& $147.848$& -    & 26 & 3 \\
050814            & $5.30$  & B & $7.44_{-1.21}^{+1.25}$ & $>227$      & $8.712$  &       -       & $3.976$  & $13.256$ & -    & 27 & 3 \\
050824            & $0.83$  & B & $4.55_{-1.20}^{+1.28}$ & $<23$       & $53.128$ &       -       & $47.816$ & $54.280$ & -    & 28 & 3 \\
050826            & $0.297$ & B & $6.95_{-1.32}^{+1.38}$ & $>140$      & $1.328$  &       -       & $-0.208$ & $2.608$  & -    & 29 & 3 \\
050908            & $3.3437$& B & $1.68_{-0.32}^{+0.21}$ & $<226$      & $2.080$  &       -       & $1.760$  & $3.104$  & -    & 30 & 3 \\
050922C$^{\rm f}$ & $2.198$ & B & $0.43_{-0.04}^{+0.02}$ & $417_{-54}^{+102}$& $-0.072$ & $510\pm40$ & $0.696$ & $0.824$ & $640\pm40$ & 19 & 1\\
051016B           & $0.936$ & B & $1.41_{-0.23}^{+0.30}$ & $<70$       & $0.072$  &       -       & $0.456$  & $0.840$  & -   & 31 &  3 \\
051109A$^{\rm f}$ & $2.346$ & B & $3.16_{-0.40}^{+0.60}$ & $540_{-120}^{+470}$ & $0.424$ & $340\pm50$ & $0.872$ & $1.448$ & $400\pm44$ & 32 & 1\\
051111            & $1.55$  & B & $4.21_{-0.27}^{+0.17}$ & $>275$      & $-0.304$ &       -       & $-0.304$ & $0.016$  & -    & 33 & 3 \\
060115$^{\rm f}$  & $3.53$  & B & $6.32\pm0.487$         & $272_{-63}^{+68}$ & $94.896$ & $110\pm20$ & $94.640$ & $96.240$ & $111\pm11$ & 34 & 3\\
060206$^{\rm f}$  & $4.048$ & B & $0.83\pm0.07$          & $394_{-41}^{+120}$ & $2.168$  & $700\pm60$  & $2.680$  & $3.192$ & $710\pm40$ &35&3\\
060210            & $3.91$  & B & $9.19_{-0.81}^{+0.84}$ & $>353$      & $-0.040$ &       -       & $0.216$  & $0.472$  & -    & 36 & 3 \\
060223A           & $4.41$  & B & $1.36\pm0.20$          & $>216$      & $0.072$  &       -       & $-0.248$ & $0.456$  & -    & 37 & 3 \\
060418$^{\rm f}$  & $1.489$ & B & $6.27_{-0.37}^{+0.38}$ & $570\pm140$ & $27.472$ & $190\pm20$    & $27.600$ & $27.664$ &  $285\pm30$ & 38& 4 \\
060502A           & $1.51$  & B & $3.52\pm0.27$          & $>444$      & $0.112$  &       -       & $-0.848$ & $1.520$  & -    & 39 & 3 \\
060510B           & $4.90$  & B & $17.32_{-1.61}^{+1.43}$& $>360$      & $136.360$&       -       & $133.032$& $138.920$& -    & 40 & 3 \\
060512            & $0.443$ & B & $1.91_{-0.51}^{+0.45}$ & $<46$       & $3.280$  &       -       & $0.016$  & $2.768$  & -    & 41 & 3 \\
060522            & $5.11$  & B & $4.23_{-0.63}^{+0.65}$ & $>235$      & $4.392$  &       -       & $2.792$  & $5.416$  & -    & 42 & 3 \\
060526            & $3.221$ & B & $4.18_{-0.34}^{+0.45}$ & $<154$      & $0.128$  &       -       & $0.704$  & $1.024$  & -    & 19 & 3 \\
060604            & $2.68$  & B & $2.63_{-0.55}^{+0.55}$ & $<195$      & $2.288$  &       -       & $1.968$  & $3.312$  & -    & 43 & 3 \\
060605$^{\rm f}$  & $3.70$  & B & $4.38_{-0.53}^{+0.54}$ & $450_{-110}^{+180}$ & $1.680$ & $96\pm21$ & $1.680$  & $4.752$  & $91\pm13$ & 44 & 3\\
060607A           & $3.082$ & B & $5.68_{-0.37}^{+0.31}$ & $>277$      & $-0.552$ &       -       & $-0.808$ & $0.792$  & -    & 45 & 3 \\
060614$^{\rm f}$  & $0.125$ & B & $16.62_{-2.84}^{+3.42}$& $55\pm45$   & $-1.360$ & $3.1\pm2.4$   & $2.864$  & $2.928$  & $6.55\pm4.95$& 46& 5\\
060707$^{\rm f}$  & $3.425$ & B & $4.96_{-0.44}^{+0.45}$ & $301_{-22}^{+16}$ & $1.928$  & $155\pm28$ & $1.992$  & $4.040$ & $154\pm16$ & 19& 3\\
060714            & $2.711$ & B & $6.04_{-0.34}^{+0.26}$ & $<171$      & $75.344$ &       -       & $75.664$ & $76.048$ & -    & 19 & 3\\
060729            & $0.54$  & B & $14.52_{-2.07}^{+2.22}$& $<79$       & $92.888$ &       -       & $91.928$ & $93.976$ & -    & 47 & 3\\
060814$^{\rm f}$  & $0.84$  & B & $11.68\pm1.09$         & $470_{-70}^{140}$ & $15.368$ & $60.9\pm4.3$ & $15.560$ & $15.880$& $62.5\pm3.5$& 48&6\\
060904B$^{\rm f}$ & $0.703$ & B & $3.42_{-0.52}^{+0.48}$ & $135_{-31}^{+64}$ & $1.104$  & $54\pm43$  & $1.680$  & $2.128$ & $61\pm48$  & 49 & 3\\
060908$^{\rm f}$  & $2.43$  & B & $1.75_{-0.13}^{+0.08}$ & $545_{-100}^{+220}$ & $0.992$ & $300\pm30$ & $1.184$ & $1.440$ & $390\pm30$ & 50 & 3\\
060912A           & $0.937$ & B & $0.59_{-0.05}^{+0.08}$ & $>205$      & $-0.032$ &       -       & $0.352$  & $0.608$  & -    & 51 & 3\\
060926            & $3.208$ & B & $0.73\pm0.09$          & $<122$      & $0.352$  &       -       & $0.480$  & $0.928$  & -    & 52 & 3\\
060927$^{\rm f}$  & $5.60$  & B & $0.67\pm0.08$   & $400_{-60}^{+110}$ & $0.168$  & $2170\pm430$  & $0.808$  & $1.192$  & $2440\pm470$& 53 & 3\\
061007$^{\rm f}$  & $1.262$ & B & $7.38_{-0.26}^{+0.24}$ & $900_{-40}^{+120}$ & $45.176$ & $1080\pm40$ & $38.456$ & $38.520$ & $1460\pm30$ & 54 &7\\
061110A           & $0.757$ & B & $8.29_{-1.03}^{+1.34}$ & $>141$      & $9.720$  &       -       & $-0.968$ & $3.128$  & -    & 55 & 3\\
061110B           & $3.44$  & B & $5.34_{-0.74}^{+0.56}$ & $>551$      & $-7.872$ &       -       & $-16.128$& $-15.424$& -    & 56 & 3\\
061121$^{\rm f}$  & $1.314$ & B & $2.21_{-0.11}^{+0.12}$ & $1400_{-170}^{+210}$ &$74.456$&$1700\pm250$ & $74.840$ & $74.904$&$2040\pm280$& 57& 3\\
061126$^{\rm f}$  & $1.1588$& B & $2.75_{-0.30}^{+0.20}$ & $1337\pm410$ & $6.552$  & $409\pm9$     & $6.680$  & $6.936$  & $446\pm18$ & 58 & 8\\
061222B           & $3.355$ & B & $3.54_{-0.41}^{+0.28}$ & $<200$      & $59.048$ &       -       & $45.672$ & $47.208$ & -    & 59 & 3\\
070110            & $2.352$ & B & $6.41_{-0.36}^{+0.47}$ & $>285$      & $-0.800$ &       -       & $-0.480$ & $1.760$  & -    & 60 & 3\\
070208            & $1.165$ & B & $2.79\pm0.59$          & $<197$      & $-0.312$ &       -       & $-0.248$ & $0.840$  & -    & 61 & 3\\
070318            & $0.836$ & B & $5.69_{-0.50}^{+0.60}$ & $>224$      & $1.168$  &       -       & $1.360$  & $2.192$  & -    & 62 & 3\\
070411            & $2.954$ & B & $8.82\pm0.47$          & $>482$      & $70.176$ &       -       & $69.856$ & $71.008$ & -    & 63 & 3\\
070419A           & $0.97$  & B & $20.86_{-3.23}^{+4.06}$& $<65$       & $-1.304$ &       -       & $18.600$ & $35.368$ & -    & 64 & 3\\
070506$^{\rm g}$  & $2.31$  & B & $0.52\pm0.08$          & $162\pm50$  & $6.312$  & $48.5\pm3.5$  & $6.376$  & $6.824$  & $57\pm5$  & 65 & 3\\
070529            & $2.5$   & B & $7.39_{-1.26}^{+1.86}$ & $>340$      & $2.008$  &       -       & $2.392$  & $2.904$  & -     & 66 & 3\\
070611$^{\rm g}$  & $2.04$  & B & $1.67\pm0.33$          & $188\pm49$  & $2.336$  & $35\pm5$      & $1.568$  & $3.744$  & $18\pm2$  & 67 & 3\\
070612A           & $0.617$ & B & $39.38_{-3.69}^{+4.62}$& $>136$      & $9.704$  &       -       & $6.248$  & $12.520$ & -        & 68 & 3\\
070721B           & $3.626$ & B & $5.06_{-0.66}^{+0.68}$ & $>624$      & $0.136$  &       -       & $0.584$  & $2.120$  & -        & 69 & 3\\
070802            & $2.45$  & B & $2.27\pm0.46$          & $>138$      & $6.120$  &       -       & $5.672$  & $10.088$ & -        & 70& 3\\
070810A$^{\rm g}$ & $2.17$  & B & $0.93\pm0.16$          & $130\pm13$  & $-0.136$ &  $65\pm4$     & $0.120$  & $0.760$  & $66\pm6$  & 71 & 3\\
071010A           & $0.98$  & B & $1.24_{-0.45}^{+0.46}$ & $<83$       & $0.992$  &       -       & $-1.056$ & $4.128$  & -        & 72 & 3\\
071010B$^{\rm f}$ & $0.947$ & B & $2.14_{-0.20}^{+0.21}$ & $101\pm20$  & $1.432$  & $36.8\pm0.8$  & $1.944$  & $2.456$  & $37.2\pm1.0$ & 73& 9\\
071020$^{\rm f}$  & $2.145$ & B & $0.51\pm0.06$          & $1010\pm160$& $-0.336$ & $1265\pm25$   & $0.240$  & $0.368$  & $1510\pm60$  & 74& 10\\
071031            & $2.692$ & B & $10.53\pm1.27$         & $<100$      & $2.880$  &       -       & $2.624$  & $5.248$  & -         & 75& 3\\
071117$^{\rm f}$  & $1.331$ & B & $0.53_{-0.05}^{+0.03}$ & $647\pm226$ & $0.016$  & $206.5\pm6.5$ & $0.464$  & $0.656$  & $231\pm13$   & 76& 11\\
071122            & $1.14$  & B & $8.94_{-2.12}^{+2.15}$ & $<96$       & $11.816$ &       -       & $-10.520$& $13.352$ & -       & 77 & 3\\
080210            & $2.641$ & B & $3.21_{-0.30}^{+0.59}$ & $>266$      & $4.448$  &       -       & $7.072$  & $9.184$  & -          & 78& 3\\
080310            & $2.43$  & B & $11.68\pm1.18$         & $<117$      & $1.152$  &       -       &  $1.280$ &  $2.176$ & -          & 79& 3\\
080319B$^{\rm f}$ & $0.937$ & B & $8.23_{-0.45}^{+0.48}$ & $1261\pm65$ & $16.848$ & $672.5\pm6.5$ & $12.420$ & $12.436$ & $1190\pm60$ & 80& 12\\
080319C$^{\rm f}$ & $1.95$  & B & $1.78_{-0.13}^{+0.07}$ & $910\pm270$ & $0.128$  & $440\pm20$    &  $0.256$ &  $0.512$ & $490\pm30$  & 81& 13\\
080330            & $1.51$  & B & $1.92_{-0.30}^{+0.44}$ & $<88$       & $0.128$  &       -       &  $0.384$ &  $0.832$ & -         & 82 & 3\\
080411$^{\rm f}$  & $1.03$  & B & $2.58_{-0.19}^{+0.21}$ & $524\pm70$  & $40.448$ & $553.5\pm5.5$ & $40.960$ & $41.088$ & $595\pm13$ & 83 & 14\\
080413A$^{\rm f}$ & $2.433$ & B & $1.74\pm0.20$          & $650\pm210$ & $1.624$  & $564\pm16$    &  $1.688$ &  $2.200$ & $570\pm20$ & 84 & 3\\
080413B$^{\rm f}$ & $1.10$  & B & $0.50_{-0.04}^{+0.06}$ & $150\pm30$  & $-0.224$ & $185\pm5$     &  $0.224$ &  $0.480$ & $200\pm10$ & 85 & 3\\
\hline
\enddata
\begin{list}{}{}
 \item [$\rm ^{a}$] Instrument: G (GRBM), B(BAT)
 \item [$\rm ^{b}$] Times of the spectrum accumulated around the peak.
They are given with reference to the GRBM (BAT) trigger time of each {\em BeppoSAX} ({\em Swift}) GRB.
 \item [$\rm ^{c}$] Peak bolometric isotropic equivalent luminosity in $10^{50}$~erg~s$^{-1}$
in the rest frame; $H_0 = 71$ km s$^{-1}$ Mpc$^{-1}$, $\Omega_m = 0.27$, and
$\Omega_{\Lambda} = 0.73$.
\item[$^{\rm d}$]References for the redshift measurements: 
(1) \cite{Djorgovski99}, (2) \cite{Metzger97}, (3) \cite{Kulkarni98}, (4) \cite{Tinney98},
(5) \cite{Djorgovski98}, (6) \cite{Kulkarni99}, (7) \cite{Bloom03}, (8) \cite{Beuermann99},
(9) \cite{Amati00}, (10) \cite{Lefloch02},  (11) \cite{Galama99}, (12) \cite{Vreeswijk99},
(13) \cite{Piro02}, (14) \cite{Garnavich01}, (15) \cite{Djorgovski01}, (16) \cite{Berger05a},
(17) \cite{BergerShin06}, (18) \cite{KelsonBerger05}, (19) \cite{Jakobsson06a},
(20) \cite{Fynbo05a}, (21) \cite{Cenko05}, (22) \cite{Berger05b}, (23) \cite{Foley05},
(24) \cite{BergerBecker05}, (25) \cite{Chen05}, (26) \cite{Bloom05}, (27) \cite{Jakobsson06b},
(28) \cite{Fynbo05b}, (29) \cite{HalpernMirabal06}, (30) \cite{Fugazza05},
(31) \cite{Soderberg05}, (32) \cite{Quimby05}, (33) \cite{Hill05},
(34) \cite{Piranomonte06}, (35) \cite{Fynbo06a}, (36) \cite{Cucchiara06a},
(37) \cite{Berger06a}, (38) \cite{Dupree06}, (39) \cite{Cucchiara06b}, (40) \cite{Price06},
(41) \cite{Bloom06a}, (42) \cite{Cenko06},
(43) \cite{Castrotirado06}, (44) \cite{Still06}, (45) \cite{Ledoux06}, (46) \cite{Fugazza06a},
(47) \cite{Thoene06a}, (48) \cite{Thoene07a},
(49) \cite{Fugazza06b}, (50) \cite{Rol06}, (51) \cite{Levan07}, (52) \cite{Delia06},
(53) \cite{Fynbo06b}, (54) \cite{Jakobsson06c}, (55) \cite{Thoene06b}, (56) \cite{Fynbo06c},
(57) \cite{Bloom06b}, (58) \cite{Perley08}, (59) \cite{Berger06b}, (60) \cite{Jaunsen07a},
(61) \cite{Cucchiara07a}, (62) \cite{Jaunsen07b}, (63) \cite{Jakobsson07a}, (64) \cite{Cenko07a},
(65) \cite{Thoene07b}, (66) \cite{Berger07}, (67) \cite{Thoene07c}, (68) \cite{Cenko07b},
(69) \cite{Malesani07}, (70) \cite{Prochaska07a}, (71) \cite{Thoene07d}, (72)  \cite{Prochaska07b},
(73) \cite{Cenko07c}, (74) \cite{Jakobsson07b}, (75) \cite{Ledoux07}, (76) \cite{Jakobsson07c},
(77) \cite{Cucchiara07b}, (78) \cite{Jakobsson08}, (79) \cite{Prochaska08a}, (80) \cite{Vreeswijk08a}
(81) \cite{Wiersema08}, (82) \cite{Malesani08}, (83) \cite{Thoene08a}, (84) \cite{Thoene08b},
(85) \cite{Vreeswijk08b}.
\item[$^{\rm e}$]References for the $E_{\rm p,i}$ measurements:
(1) \cite{Amati06}, (2) \cite{Ulanov2005}, (3) This work, (4) \cite{Golenetskii06a}
(5) \cite{Amati07}, (6) \cite{Golenetskii06b}, (7) \cite{Mundell07}, (8) \cite{Perley08},
(9) \cite{Golenetskii07a}, (10) \cite{Golenetskii07b}, (11) \cite{Golenetskii07c},
(12) \cite{Golenetskii08a}, (13) \cite{Golenetskii08b}, (14) \cite{Golenetskii08c}.
\item[$^{\rm f}$] GRBs with firm measurements of $E_{\rm p,i}$, $T_{0.45}$ and $L_{\rm p,iso}$,
used to derive the best-fitting parameters of the correlation.
\item[$^{\rm g}$] From the BAT data we could constrain only $E_{\rm p,i}$, while no information
on $\alpha$ and $\beta$ could be derived. These values for $E_{\rm p,i}$ are not confirmed by
\cite{Sakamoto07}. These GRBs were not included in the sample used to fit the correlation, but
just displayed in Fig.~\ref{f:Ep-Lp2-T045-lim}. Their peak luminosities were computed assuming
$\alpha=-1$ and $\beta=-2.3$.
\end{list}
\end{deluxetable}

\twocolumn

%
%
\begin{table*}
\centering
\caption{Best-fitting parameters of the correlation between $E_{\rm p,i}$, $L_{\rm p,iso}$
and $T_{0.45}$.}
\begin{tabular}{l|l|c|c|c|c|c|c}
\hline
Sample & Correlation & $a$ & $b$ &  $q$  & $\sigma_{log(E_{\rm p,i})}$ & $\chi^2/{\rm dof}$ & Prob.\\
\hline
all (40) & $E_{\rm p,i} \propto L_{\rm p,1s}^a\ T_{0.45}^b$ & $0.49\pm0.07$ & $0.35\pm0.12$ & $1.28_{-0.20}^{+0.19}$ & $0.15_{-0.03}^{+0.05}$ & $40.8/36$ & $0.27$ \\
all (40) & $E_{\rm p,i} \propto L_{\rm p,var}^a\ T_{0.45}^b$ & $0.46\pm0.07$ & $0.35\pm0.13$ & $1.32\pm0.20$ & $0.17\pm0.04$ & $40.0/36$ & $0.30$ \\\hline
{\em Swift} (27) & $E_{\rm p,i} \propto L_{\rm p,1s}^a\ T_{0.45}^b$ & $0.56_{-0.11}^{+0.10}$ & $0.39_{-0.16}^{+0.15}$ & $1.12_{-0.30}^{+0.28}$ & $0.17\pm0.04$ & $27.1/23$ & $0.25$ \\
{\em Swift} (27) & $E_{\rm p,i} \propto L_{\rm p,var}^a\ T_{0.45}^b$ & $0.52_{-0.10}^{+0.11}$ & $0.38_{-0.17}^{+0.16}$ & $1.16\pm0.29$ & $0.18_{-0.04}^{+0.07}$ & $27.5/23$ & $0.24$ \\
\hline
all (40) & $E_{\rm p,i} \propto L_{\rm p,1s}^a\ T_{0.45}^b$ & $0.55\pm0.02$ & $0.50_{-0.03}^{+0.02}$   & $1.03\pm0.05$     & $[0]$ & $218/37$ & --\\
all (40) & $E_{\rm p,i} \propto L_{\rm p,var}^a\ T_{0.45}^b$ & $0.51\pm0.02$ & $0.485\pm0.025$  & $1.07_{-0.04}^{+0.05}$  & $[0]$ & $260/37$ & --\\
\hline
\hline
\end{tabular}
\label{t:parametri_ep_LT}
\end{table*}
%

\section*{Acknowledgments}
We acknowledge the anonymous referee for useful comments.
This work is supported by ASI grant I/011/07/0 and by the
Ministry of University and Research of Italy
(PRIN 2005025417).
We gratefully acknowledge the contributions of dozens of
members of the BAT team who built and maintain this instrument.


\bsp

\label{lastpage}

\end{document}